\documentclass{ws-ijmpeQU}

\begin{document}

\catchline{}{}{}{}{}
 \title{QUARKS IN THE UNIVERSE\footnote{Dedicated  to
Walter Greiner on occasion of his 70th birthday.}}

\author{\footnotesize JOHANN RAFELSKI}

\address{Department of Physics, University of Arizona\\
Tucson, AZ 85718, USA}

 \maketitle
\thispagestyle{empty}

\centerline{\small September 15, 2006}


\begin{abstract}
 
\end{abstract}

\section{Local change in the vacuum structure: from QED to QCD}
Matter in its present  form  was formed when our
 Universe emerged from the quark-gluon phase (QGP)  
at about 30$\mu$s into its evolution. To explore this early period
in the laboratory,  we study highly excited matter formed in 
  relativistic heavy ion collision experiments:
heavy nuclei crash into each other, and form compressed and energetically 
excited nuclear matter,  resembling in its key features 
the stuff which filled the early Universe.
 In these experiments we further explore  the physics 
of the vacuum structure of strongly interacting gauge theory, 
Quantum Chromodynamics (QCD). The {\em common} beginning for both,  
heavy ion collisions, and vacuum structure investigations, is the  physics of 
the quantum electrodynamic (QED) vacuum in the presence of the supercritical 
external field  that is formed when two highly charged  heavy ions are 
brought together near  to  the Coulomb barrier in a considerably lower reaction 
energy  collision.\cite{GMR} 

QED of strong fields research program was initiated by Walter Greiner in Frankfurt
many years ago.  The 2nd island of super-heavy elements  at $Z=164$ demanded 
the understanding of relativistic atomic physics structure, and more specifically, 
the understanding of what happens when the most tightly bound  electrons 
disappear into the  lower Dirac continuum.
 At the Montreal meeting  in late August 1969,\cite{GreinerMontreal} 
shortly following his first study of atomic structure of super-heavy 
elements,\cite{pieper}  Walter Greiner made the following comments:  
{\it     Heavy-ion physics is the tool \ldots.
\ldots we find in the future elements in the area of $Z=164$ \ldots with accelerators. 
\ldots this would lead us into new field of quantum electrodynamics \ldots 
of strong fields \ldots  an unsettled problem and a new and rich field of research.
\ldots if you come to very high $Z$-numbers the 1s-electron levels \ldots dives into the
lower continuum.} Walter prophecy came true with regard to his theory of high  $Z$-atoms.
As for ``future elements in the area of $Z=164$, '' this remains today good  material for 
science fiction. 

Along with a few  other  young students
I   joined this adventure in a new field,  the QED of strong fields. 
One Saturday morning, in the Fall of 1971,  
 the process of  positron auto-ionization   was finally understood.\cite{Muller:1972zz} 
After a bit more work the charged vacuum  was born:\cite{Rafelski:1974rh}  the positron
carried out the positive charge, while the 
balancing negative charge was in the vacuum,  localized near the source of the supercritical field. 
An important consequence of all this was the proper understanding of the behavior of 
electrons and positrons  in rapidly changing strong fields, from which the prediction of the shape of 
the emitted positron spectra emerged.  

Several  technically challenging    experiments followed, yet the detection of
the spontanouse positron production in  supercritical fields 
remains an open subject, despite  years of diligent work. The QED
vacuum kept its secret, buried in the high noise generated by other processes
accompanying atomic and nuclear reactions of very heavy atoms.
We have  not  demonstrated 
that the QED vacuum state can change locally. 

We now search for another local vacuum modification,  
the melting of the structure of the 
vacuum of strong interactions. Unlike the case of QED, where the local 
non-perturbative structure  is created in the experiment, we aim here to 
locally dissolve  the global non-perturbative, color charge structure,
and to locally liberate quarks  confined in hadronic particles.
In both cases, the QED of strong fields, and the QCD at high temperature, 
we probe the same principle, that a local change of the vacuum 
state is possible.   

By showing  that it is possible to dissolve color confining vacuum structure,
we demonstrate that the vacuum state can change locally.   We 
further complete the 
understanding  of the origin of the mass of matter, which is as we 
believe today, due to the confinement properties inherent in the 
non-perturbative nature  of the true QCD vacuum.  We 
further learn how the QGP energy becomes matter in the process we call hadronization.
We thus learn about the matter formation mechanisms   in the  expanding   Universe.
There   hadronization occurs at  about $T_h=160$\,MeV .

The research area of 
high energy heavy ion collisions emerges as a new field in the '70-s.
 The first  application  is  at that time the exploration of compressed
nuclear matter.   Walter Greiner is among the first to propose  
hydrodynamic description of the evolution of the
strongly interacting matter. He proposes an interpretation 
of some results in terms of shock waves.\cite{Baumgardt:1975qv} 
These could help compress 
nuclear matter to conditions expected in the interior of 
the neutron stars.\cite{Stoecker:1982qx}  

Our understanding of the hot hadronic matter
formed in these reactions expands  rapidly. We recognize that already at 
rather modest heavy ion  reaction energies we can encounter 
deconfinement.\cite{Hagedorn:1980kb,Rafelski:1980rk,Rafelski:1982rq,Rafelski:1983ja}  
By 1982 our theoretical work   suggests
that QGP is formed in heavy ion collisions and can be observed. 
 
Two proposed dedicated experimental facilities, the accelerator projects 
at the  LBL (Venus) and GSI (SIS100), do not attract funding  in the early eighties.
Despite this initial setback the field of nuclear physics  moves decisively into this new area.
The interest in  the  QGP research program grows rapidly,  both in Europe and USA.
The theoretical effort is soon supported by experiment, 
with a large number of experimental  nuclear physicists   entering into 
research collaborations with particle physicists, and jointly 
developing full fledged experiments  at the 
particle physics laboratories, at CERN in Geneva, and at BNL in New York.

\section{Creation of Matter in Laboratory} 
\begin{figure}[th]
\centerline{\psfig{file=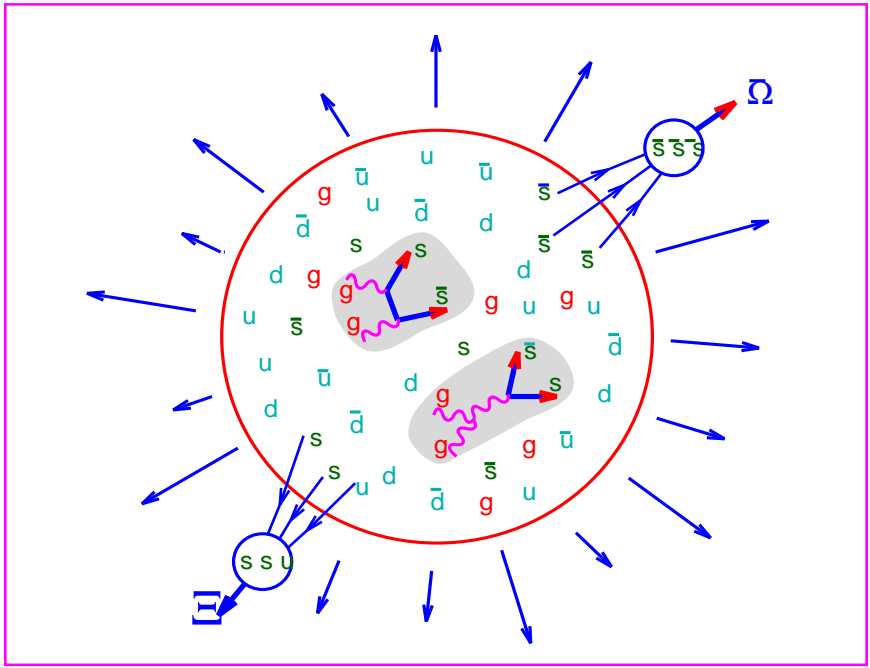,width=7.8cm,angle=-90}}
\vspace*{8pt}
\caption{A schematic illustration by example of the two step particle formation process:
within QGP strangeness is produced in gluon fusion $GG\to s\bar s$,  
and later combined into final state particles.\label{twostep}
}
\end{figure}

In  laboratory experiments,
there are two primary steps in the particle production from 
QGP as illustrated in figure \ref{twostep}:\begin{itemlist}
\item
cooking of the energy content  towards
 QGP  $u,d,s$ quarks and  $G$ gluon yield (chemical) equilibrium;
\item
combination of quark content into final state hadrons, in figure \ref{twostep},
the precooked strangeness
content combines into  $\overline\Omega(\bar s \bar s \bar s)$
and  $\Xi^0(ssu)$ .
\end{itemlist}

The  hadronic particles  emerge in a quark combination process;
see arrows in   figure \ref{twostep}. The
hadronic particle can be surface-produced in this way 
during the entire history of plasma evolution.\cite{Danos:1982ar,Grassi:1994ng}
However,  given the high collective outflow velocity of the matter, driven by the 
high collisional compression pressure,  it is widely believed
that the bulk of hadrons emerges at the end of the expansion 
in the global volume dissociation. At this point in time,
 across the fireball volume, the 
temperature decreases to the point at which  
the deconfined phase cannot continue to exist. This final 
breakup of the QGP phase formed in the laboratory is, as
pion interferometry HBT results show,  a very fast process.  
We thus conclude that   the dense matter fireball
 expands  in an explosive manner,  and undergoes a fast bulk ``hadronization''.

 The resulting particle yields are well described by the   statistical hadronization model  (SHM).
SHM originates in the Fermi-hypothesis: strong interactions  
saturate the   quantum particle production
matrix elements. Therefore, pursuant to the  golden rule of quantum mechanics (attributed to
Fermi) the yield of   particles  is given 
by the accessible phase space. For a more detailed discussion of 
SHM and its parameters we refer to the recent review. \cite{Torrieri:2004zz} 

In the original Fermi model the accessible 
phase space is considered in terms of the available energy. Today we refer to this method as
micro-canonical. Normally, we use the grand canonical approach,
which substitutes for the  total  available fireball energy 
a  temperature-like parameter $T_f$.  Even if $T_f$ is reported
in context of SHM, that does not
mean by necessity  that there is an equilibrated gas of hadrons. 
Particle spectra and abundances may imply different values of $T_f$ 
if following the thadron formation there is a period in which hadrons interact 
and reequilibrate. In this case, $T_f$ is called chemical freeze-out 
(particle formation) temperature, and another parameter $T_t$ appears, 
the thermal freeze-out temperature. If QGP is the source of hadrons,   $T_f$ is 
closely related to the hadron source temperature $T_h$  of the QGP .

The state of the art of SHM 
 is  today more complex than in time of Fermi.
To make a quantitative model we  must deal with strong
interactions  among particle , this is done   introducing the production of
hadron resonances. In addition we consider   
chemical potentials associated with all conserved quantum numbers.
We can either work with baryon number  $B$, hyperon charge $Y$ and electrical charge $Q/|e|$,  or
 the net number $N_i$ of each of the 
three valance quark flavors,
\begin{equation}
N_i=q_i-\bar q_i, \quad  i=u,d,s 
\end{equation}
Specifically, 
\begin{equation}
B=\frac 1 3(N_u+N_d+N_s), \quad 
Q=\frac 2  3 N_u-\frac 1 3 N_d-\frac 1 3 N_s, \quad 
Y=B+S=B - N_s
\end{equation}
In general there are three chemical fugacities $\lambda_i$, or 
potentials, $\mu_i=T\ln \lambda_i$, with either 
$i=u,d,s$ or equivalently $i=b,Q,Y$ or any linearly independent combination
of the three which must be introduced to be able to satisfy the conservation 
of these three quantities. One often refers to light quarks
by $\lambda_q\equiv \sqrt{\lambda_u\lambda_d}$, $\mu_q=(\mu_u+\mu_d)/2$. 
 
Conservation laws do not tell us anything about actual  `filling' of phase
space.  For example in laboratory experiments,  initially there are very few,
 if any,   strange quark pairs present.
As the collision reaction progresses, the yield of strangeness grows.
This is described by a parameter $\gamma_s(t)$ which expresses how
close one is to a yield expected when the system had long time to cook 
strangeness in the QGP.

Within the usual framework of 
statistical thermodynamics of quarks and gluons   these 
parameters enter the quantum Fermi and Bose distributions, for example
for the conserved strangeness flavor we have 
\begin{equation}\label{sdens}
s=\int \frac{d^3p}{(2\pi)^3} \frac 1 
   {\lambda_s^{-1}\gamma_s^{-1}e^{E(p)/T}+1} \qquad 
\bar s=\int \frac{d^3p}{(2\pi)^3} \frac 1 
   {\lambda_s\gamma_s^{-1}e^{E(p)/T}+1} 
\end{equation}
In a good approximation,
\begin{equation}
\gamma_s\simeq \frac{s+\bar s}{s_{\rm eq}+\bar s_{\rm eq}}
\end{equation}
and similarly for all other   quark flavors.  Here the equilibrium distribution 
arises for $\gamma_s\to 1$. One often refers to light quarks
by $\gamma_q\equiv \sqrt{\gamma_u\gamma_d}$. 

It is not customary   to introduce
a chemical potential associated with the fugacity $\gamma_i$ since in any
physical system,  as reaction time evolves,  $\gamma_i(t)\to  1$ 
maximizing the entropy . Thus the associated chemical potential is not time independent, but
rapidly evolves to zero. On the other
hand,  the conserved quantum number chemical potentials 
\begin{equation}
\mu_{B}=\frac 3 2 (\mu_u+\mu_d),\quad  
\mu_Q=\frac23\mu_u-\frac13 \mu_d -\frac13\mu_s,
\quad   \mu_Y=\mu_B-\mu_s
\end{equation}
are normally constant in time. However, in the heavy ion reaction environment 
the expansion-driven cooling of the  system leads to a decreasing value of $T$. 
Note that the dimensionless quantity $\mu_B/T$ is nearly conserved in hydrodynamic
expansion of QGP. 

\section{In search of a new phases of matter}
\subsection{Phase boundary}
The best way to discover a new phase of matter is to 
find a phase boundary. We  recall  Gibbs definition of the 
phase boundary. For the case of  chemical equilibrium:\\
a) The pressure  is equal, or else mechanical force would move the boundary 
in space, or in time,  until the pressures in both domains are equal; \\
b) The temperature is equal, or else radiative processes would transport
energy between domains in space until this condition is reached;\\
c) The chemical potential(s) is(are) equal, or else particle transport across boundary 
would change the particle number so that this condition is satisfied. \\
The last condition follows from the first law of thermodynamics. 

However, unlike the Gibbs case, we deal here with systems in which  particle number is not conserved,
so we need to modify this condition to get:\\
c') The `conserved quantum number' chemical potential(s) is(are)  
equal, or else particle and antiparticle transport across boundary 
would change the baryon number, charge etc, so that this condition is satisfied. 

If the particle yield equilibrium (chemical equilibrium) 
cannot be assured, for example when the physical constraints are evolving fast, 
we speak of chemical non-equilibrium. In this case  the 
phase boundary is defined by   micro-canonical properties which have not 
otherwise been considered:\\
d) The near-conservation of entropy, and hard-to-form particles across the 
phase boundary. \\
The entropy cannot decrease at the phase boundary, but it could increase. However 
since QGP is an entropy-rich phase undergoing a fast phase transformation, this 
should occur  without significant on the scale of entropy already generated, entropy enhancement. 
There are several options to accomplish this, {\it e.g.\/} by volume expansion or/and 
phase space occupancy $\gamma_i\ne 1$.
 
In chemical non-equilibrium the Gibbs condition c) and its variant c') 
have to be applied to both the particle and antiparticle number 
separately. Using strangeness as  example, see Eq.\,(\ref{sdens}),  we write:
\begin{eqnarray}
&&{\rm for}\ s{\rm-quarks} \ \mu^t_s =T\ln (\gamma_s\lambda_s)=T\ln \gamma_s+\mu_s \\
&&{\rm for}\ \bar s{\rm-quarks}\  \mu^t_{\bar s} =T\ln (\gamma_s\lambda_s^{-1} )=T\ln \gamma_s-\mu_s
\end{eqnarray} 
Note that  $\mu^t_s-\mu^t_{\bar s} =2\mu_s$ is independent of $\gamma_s$, and 
assures that net strangeness is conserved. Similarly,  $\mu^t_s+\mu^t_{\bar s} =2T\ln \gamma_s$
is independent of $\mu_s$ and assures that number of strange quark pairs (up to additive constant
and factor two) is conserved. 

\subsection{Non-equilibrium phase boundary in heavy ion reactions}
An important difference between $\gamma_i$ and $\mu_i$ is that the Gibbs
criteria of a smooth functional 
connection across phase boundary do not apply to $\gamma_i$. The Gibbs condition,
that at the phase  boundary transport of particles of given conserved number content
must vanish {\em requires} in general for continuous $\mu_i$ 
 a discontinuity in all $\gamma_i$, since in general the size of the phase space is not 
the same in the  two matter phases considered. For example, conserving entropy in hadronization
amounts to hadronization of equilibrated QGP into oversaturated HG:
\begin{equation}
\gamma_q^{\rm QGP}=1\to \gamma_q^{HG}\simeq  e^{m_\pi/2T}\simeq 1.6.
\end{equation} 
Similarly, since strangeness phase space size is about 2.5 times smaller in HG compared
to QGP, chemically equilibrated strange QGP  implies $\gamma_s^{\rm HG}\simeq 2.5$\,.

We note further physical meaning of the non-equilibrium parameters with regard to hadron 
yields. Comparing the 
yield of strange to non-strange hadrons of the same type ({\it e.g.\/} nucleon with a hyperon,
kaon with a pion etc.) we are evaluating the ratio
$\gamma_s/\gamma_{u,d}$ Similarly, the relative yield  of baryons ($\propto \gamma^3_q$) 
to mesons ($\propto \gamma^2_q$) is  controlled at fixed values of 
$\gamma_s/\gamma_q, T$  by 
$\gamma_q$.  The observed baryon-to-meson ratio in nuclear collisions at RHIC
is strongly enhanced compared to the yield seen in $pp$ reactions at the same energy.
Thus we know that the mechanism of baryon production in heavy ion 
reactions (quark combination mechanism) 
is different from $pp$ reactions. If these yields can be 
described well by SHM model, we are expecting $\gamma_q> 1$.  

We have given here highlights of what needs to be 
considered in setting up the QGP fireball 
breakup into hadrons within the SHM. More details can be found in the 
manuals to SHARE suite of programs (Statistical HAdronization
with REsonances).\cite{Torrieri:2004zz} Further discussion
of the impact of the heavy ion dynamics on phase boundary
have also been recently described.\cite{Rafelski:2005md}

\section{Strangeness and the Discovery of QGP}
\subsection{Measuring QGP degrees of freedom}\label{str}
The   QGP at hadronization,  in the early Universe as in the laboratory,
 consists of $u,d,s$ and their
anti-quarks $\bar u,\bar d,\bar s$. In laboratory 
experiments, strangeness formation continues throughout  the temporal evolution
of the plasma until the break up temperature $T_h$. Because of the coincidence of
scales with $T_h\simeq 170\pm 20$ MeV and $2m_s\simeq 190\pm30$ MeV 
being not very different,  
 the yield of strangeness  is a natural probe of QGP. 

One way to understand if QGP has been formed is to 
study the available number of degrees of freedom. This
can be accomplished by comparing the strangeness pair yield $N_s$ with entropy $S$.
We  denote here the yield of strange quarks by $N_s$, 
which is the  same as the yield of strange quark pairs.
Both $N_s$ and $S$ are extensive in the volume and thus not subject to 
dependence on precise collision history. Their ratio is 
in effect, up to a factor 4,  the ratio of strangeness 
degeneracy $g_s$ to all active degrees of freedom in plasma $g_{\rm QGP}$.
The factor 4 allows  (in good approximation) for
the entropy per particle content in a nearly massless gas:
\begin{equation}
\left.\frac{N_s}{S}\right|_{\rm QGP}=\frac{{3_c 2_s \cdot \tilde\gamma_s}}{ 
        {[(2+\tilde\gamma_s)_f3_c2_s+8_c2_s]4}} \simeq 0.03,
\end{equation}
Here $\tilde\gamma_s<\gamma_s$ allows for the reduction in the effectively
acting massless strangeness degrees of freedom due to the strange quark mass, $m_s\ne 0$,
and due to under-saturation of the phase space described by factor $\gamma_s$.

The final value,
\begin{equation}
\gamma_s(t_h)\equiv \gamma_s^{\rm QGP}(\sqrt{s_{\rm NN} },A) 
\end{equation}
reached at time of hadronization $t_h$, is growing with increasing collision 
energy $\sqrt{s_{\rm NN}}$, and with increasing participant number $A$,
{\it i.e.\/} volume $V\propto A$.
 Thus  we expect  that as function of these
variables, $\gamma_s^{\rm QGP}\to 1$ 
for a sufficiently large $\sqrt{s_{\rm NN}},A$. In this   limit the 
analysis of experimental data would yield $N_s/S\to 0.03$.  

The measurement
of this value for $N_s/S$, which is believed to be preserved in hadronization
of QGP, amounts to a  measurement of the relative strength of strangeness
among all QCD degrees of freedom. An in-depth analysis of the experimental
conditions shows that for the most central RHIC collisions we indeed have
 $N_s/S\to 0.03$. \cite{Rafelski:2004dp}

\subsection{Strange antibaryons}
A promising indicator for the formation of QGP is 
the anomalous yield of strange antibaryons. \cite{Rafelski:1980rk} 
 Their production occurs through combination of earlier
produced quarks and thus anti-strangeness rich 
QGP is a particularly good source of otherwise 
more rarely produced strange antibaryons. Enhanced
production of $\overline\Lambda(\bar s \bar q \bar q), 
\overline\Xi(\bar s \bar s \bar q), \overline\Omega(\bar s \bar s \bar s)$
increasing with the $\bar s$ content is the   signature of QGP.

The detection of these particles  is assisted by their 
natural radioactive decay patterns, which can be seen tracking
secondary charged particles. For example, to observe a 
$\overline{\Xi^-}(\bar s \bar s \bar d)$ we note its decay:
\begin{equation}
\overline{\Xi^-}(\bar s \bar s \bar d)\to 
       [\overline\Lambda(\bar u \bar d \bar s) 
                       \to \bar p+\pi^+] + \pi^- . 
\end{equation}

The simplest yield ratio to consider is $\overline\Lambda$, $\overline p$.
After cancellation of combinatorial and phase space factors
this ratio is  determined by relative quark yields
available at hadronization.  If no QGP were formed 
one could at best hope for chemical equilibrium yields in 
the hadron gas matter. In both cases, aside of directly produced 
$\overline\Lambda$, $\overline p$, there are decays
of resonances. We assume here that these multiply the 
yields of  $\overline\Lambda$, $\overline p$ by the same factors
irrespective if these are originating in QGP or HG.
For the purpose of comparing the magnitude of this ratio
originat first  ating in either QGP or HG,   these corrections can be 
ignored at first . 

In  a baryon-rich QGP environment the 
light antiquark  $\bar u,\,\bar d$ abundances are suppressed 
by the baryochemical potential, while  $\bar s$ is suppressed
by strange quark mass, and we find:
\begin{equation}
 \left.\frac{\overline\Lambda}{\overline p}\right|_{\rm QGP}=
 \frac{N_{\bar s}N_{\bar u}N_{\bar d}}{N_{\bar u}N_{\bar u}N_{\bar d}}
 \simeq \frac 1 2 \frac{m^2_s}{T_h^2}K_2(m_s/T_h)e^{(\mu_u-\mu_s)/T_h} 
=0.9 e^{(\mu_u-\mu_s)/T_h},
\end{equation}
where the last equality follows  the currently accepted value $m_s/T_h\simeq 0.7$.

The  thermal yield originating in
the hadron phase comprises, in place of strange quark mass 
suppression, the hadron phase space suppression factor:
\begin{equation}
\left.\frac{\overline\Lambda}{\overline p}\right|_{\rm HG}=
\left(\frac{m_{\overline\Lambda}}{m_{\overline p}}\right)^{3/2}
e^{-(m_{\overline\Lambda}\,-\,m_{\overline p})/T_f}e^{(\mu_u-\mu_s)/T_f}
=1.3 e^{-180\,{\rm MeV}/T_f}e^{(\mu_u-\mu_s)/T_f}
\end{equation}
For $T_f\simeq 160\pm20$  MeV we obtain a significant reduction 
of the expected relative yield, which is also clearly iless than  unity. 
If the chemical equilibrium in HG is not reached we further have
a multiplicative factor $\gamma_s/\gamma_q$. It is very hard,
indeed impossible, to ever obtain a result that would exceed unity 
in case of HG-based production. 

In  SHM fits of  ratio  $\overline\Lambda$, $\overline p$  the presence of QGP is expressed
by the magnitude of $\gamma_s/\gamma_q$, which 
in order to accommodate the large strangeness content of QGP, can exceed unity. 
The additional baryons produced by the quark combination mechanism
(comparing to HG yield)  imply that $\gamma_q>1$. We further note
that the above argument can be 
repeated for  $\overline\Xi+/\overline\Lambda$,  $\overline\Omega/\overline\Xi+$,
easily and exactly in the same way .
Thus arises the original prediction that strange antibaryon enhancement grows
with the strangeness content.

\begin{figure}[th]
\centerline{\psfig{file= 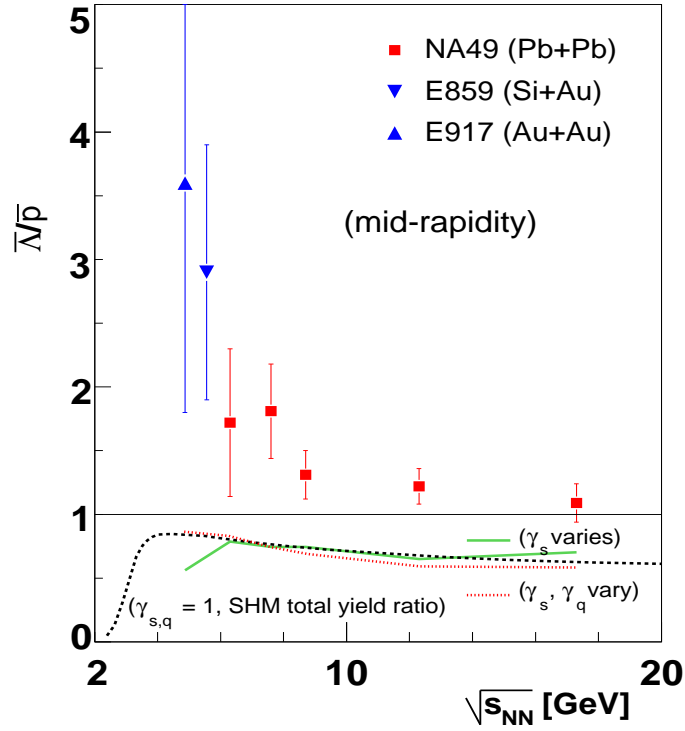,height=10.cm,width=9cm}}
\vspace*{3pt} 
\caption{
Top: observed mid rapidity  particle yield  ratio 
$\overline \Lambda/\overline p$ as function 
of nucleon-nucleon reaction energy $\sqrt{s_{\rm NN}}$. 
Bottom: statistical QGP hadronization total yield ratios in different
QGP breakup scenarios.
NA49 compilation of own, AGS data and theoretical results.
\label{aLap}
}
\vspace*{-3pt} 
\end{figure}

It is of some interest
to see     how current AGS and SPS experimental  results compare to this 
initial  prediction. The results are shown in figure \ref{aLap}, based on compilation 
of  data and theoretical results by the NA49 collaboration.\cite{Alt:2006dk} We see
that  the central rapidity ratio  $\overline\Lambda/\bar p$ is well above unity at all available 
reaction energies. With decreasing reaction energy,  
this ratio  increases. This  suggests that the  increasing baryon density 
and its suppressing effect outweigh any reduction of the relative
yield due to reduced   strangeness abundance.

At the bottom of  figure \ref{aLap}
predictions based on the SHM fits to the experimental data 
made by different groups are shown. For
the three  highest  reaction energies, the discrepancy
between theoretical interpretation and experimental data is fully accounted
for by the need to correct the experimental ratio for the included
weak decays $\overline\Xi\to \overline\Lambda$, 
and the fact that the thermal rapidity distribution of  $\overline\Lambda$ is narrower than that
of $\bar p$, which enhances the central rapidity ratio. Despite the large
error bars, it is hard to explain the trend at very low reaction energies, considering that
we would not expect that QGP is formed  below a certain energy threshold.     We will need further 
data to resolve this intriguing trend. The experimental difficulty here is the relatively 
low yield of all very massive particles, and the decreasing sensitivity of antibaryon
 detection.

\section{The Early Universe}
In order to address the physics of the early Universe 
using the results we obtain in the laboratory, 
we must extrapolate the properties of QGP  
to the conditions prevailing in the early Universe. The experimental environment 
we expect to create in relativistic heavy ion collisions
differs  from what we know about the early Universe in several ways,
which our extrapolation must bridge. In addition,  
we have to face the unknown  physical properties of dark matter and dark energy. 
We discuss these three issues before closing this presentation.

\subsection{The matter-mirror matter symmetry}
This symmetry is best described by the entropy  $S$ content  per nett
baryon $B$, that is, the excess of baryon, over antibaryon, number. 
Experiments at RHIC reach $S/B|_{\rm RHIC}=300$--400. However, our
Universe is much more symmetric at the time of hadronization. 
The study of the photon content of the Universe leads to:
\begin{equation}
\eta\equiv  \frac{B}{N_\gamma}=(6.1\pm0.15)\, 10^{-10}\,.
\end{equation}
Allowing  for the  $e+e^-$annihilation reheating 
in the late stage   the entropy content of the visible Universe is 
\begin{equation}\label{SB}
\left.\frac{ S}{B}\right|_{\rm Universe}= \frac{8.0}{\eta}=(1.3\pm0.1)\,10^{10} . 
\end{equation}
Using $S/B$ as a measure, the early Universe has been $3\cdot 10^7$ more matter symmetric
compared to RHIC, assuming here that possible decay of dark matter
has not significantly diluted the $S/B$ ratio (see below). 

Another way to understand
this difference is to look at the baryochemical potential $\mu_B$, a quantity which 
controls baryon and antibaryon density. In the RHIC experiments,
we find $\mu_{\rm B}=24\,$MeV, while in the early Universe this number 
is  $1.1$eV;\cite{Fromerth:2002wb} the corresponding ratio is similar to what we noted above, 
$2\cdot10^7$. At LHC we expect to measure  $\mu_{\rm B}\simeq 1$--2\,MeV, 
reducing the difference with the early Universe to a mere ``million''. 
On the other hand, this particular difference is not very relevant, current 
evaluation of the phase structure of matter in this domain of    $\mu_{\rm B}$  suggests 
that  the properties of QCD matter change smoothly.\cite{Philipsen:2005mj}  

 $\mu_{\rm B}$  is a very tiny fraction of the  temperature.  At hadronization in the early Universe  
the $\mu_{\rm B}/T$ ratio fixes the magnitude  of the baryon  density ,  
\begin{equation}
 {\mu_{\rm B}\over 3T}\simeq  2\cdot10^{-9}\simeq {B\over S}
    \propto {n_q-n_{\bar q}\over n_q+n_{\bar q}}
\end{equation}
 The   frequently  asked question, 
is what is the origin of this very   small value of baryochemical
potential, or equivalently net baryon number. The  absence of 
antimatter in our neighborhood forces us to think about  baryo-genesis. 
The orthodox point of view is that this small value was established well ahead of the
temperature of interest here, in terms of a  baryon-genesis mechanisms which occur
well above QCD phase transition scale of energy.\cite{Dolgov:1991fr} 

We look for baryo-genesis at high energy since we did not find any trace of it
in large laboratory experiments carried out at $T=1/30$ eV, nor did we see 
anomalous baryon number non-conserving effects in elementary collisions.  
On the other hand, the smallness of the required asymmetry creates a 
tantalizing opportunity to seek out other baryo-genesis 
effects at lower temperatures.\cite{Farrar:1993hn} 
In our opinion, it is not   possible to exclude that baryon number
violating effects occur as late as the era of QGP hadronization. 
 
One may even doubt that baryo-genesis is required at all,
and instead the baryonic Universe we see around us 
could be simply a large fluctuation, with other domains 
containing the missing antibaryons. 
In this context it is helpful to consider the magnitude of 
 fluctuations in baryon number density.
 Applying the usual formulas of grand canonical statistical physics
and recalling that since in quark phase where $T>>\mu_{\rm B}$
\begin{equation}
\overline B \propto  \mu_{\rm B} T^2 
\end{equation}
Here we indicate the grand canonical average by the over-line.
We  find the thermal fluctuation in baryon number in a given 
volume with mean baryon number  $\overline B$:
\begin{equation}
\overline{B^2}-\overline B^2=T\frac{\partial \overline B}{\partial \mu_B}
   =\frac {T }{\mu_{\rm B}} \overline B
\end{equation}

The normalized probability distribution of finding baryon number $B$   where  $\overline B$
would be expected is: 
\begin{equation}
P(B)=\frac{1}{2\pi} \frac {1}{(T/\mu_{\rm B})\overline B }\,
     e^{{{-(B-\overline B)^2}\over { \overline B 2(T/\mu_{\rm B})}}}
\end{equation}
Considering that $\mu_{\rm B}/3T\simeq 10^{-9}$, strong fluctuations in 
baryon number occurs in volumes in which $\overline B<10^9$, and thus 
considering Eq.\,(\ref{SB}),
comprising up to about $10^{19}$ particles.  One should note that the
fluctuation we have considered here  is purely thermal and 
excludes dynamical and non-equilibrium effects which 
arise, {\it e.g.\/} in hadronization of QGP. 

\subsection{Time constants}
The typical life span of  QGP  formed in the laboratory is
determined by the comparison of the 
nuclear size $R=6$\,fm to the speed of expansion, which
is about $v=0.6$c. Thus  $\tau\simeq 6/0.6$\,fm/c$\simeq 3\,10^{-23}$\,s.
This amount of time  will not allow the equilibration of particles that interact only 
by  weak, or electromagnetic interactions. In this short time 
even    some strong interaction components in 
the QGP  will not fully equilibrate. For example, computations of strangeness yield
equilibration in the deconfined QGP show that in currently available
experimental conditions only the most central RHIC collisions at maximum 
energy come to about 90\% chemical equilibration.  
The study of strangeness chemical equilibration is thus of considerable interest
and helps us understand the  QGP reaction time.  
All   heavier  flavors $c,b,t$ are, like  the weak and EM particles, practically  decoupled; their 
abundance is established in first most energetic parton collisions.

The natural time  constant in the early Universe is much longer, since the expansion 
velocity is the Hubble constant, at the time of hadronization. This follows from the 
two equations which govern the dynamics of the homogeneous Universe {\em adiabatic}
expansion. One of these  corresponds exactly to the heavy ion situation; it 
describes the   entropy conservation in the Universe expansion:
\begin{equation}\label{adexp}
{d\epsilon\over \epsilon +P}=-3 {dR\over R},
\end{equation}
where $\epsilon$ is the energy density of the gravitating matter,   $P$ its pressure,
and $R$ the radial Universe size/scale. The  expansion of the QGP phase can be described 
by a similar expression, which allows for different dynamics of longitudinal and 
transverse expansion. 

The the other, dynamical, equation is the   Friedman equation which one 
obtains for the Robertson-Walker Universe using the energy-momentum tensor of 
dust matter in the Universe:
\begin{equation}\label{T00}
\left(\frac{\dot{R}}{R} \right)^2+{k\over R^2}={8\pi G\over 3}\epsilon +{\Lambda\over 3}
\end{equation}
Here $G$ is the gravitational constant, $R$ is the size scale of the Universe, 
$\Lambda$ is the cosmological constant.
 $k$ is the curvature  index. For $k=+1$, the Universe is closed (analogous to a sphere 
in 3d) and for $k=-1$, it is open.  A flat Universe with $k=0$ (analogous to a sheet in 3d)
is  favored by observational cosmology.
The reader is invited to consult reviews and reference updates of this rapidly 
evolving field,  such as found e.g. in the PDG-biannual review volume.\cite{Yao:2006px}  

Inspecting the form of Eq.\,(\ref{T00}) we see that a natural time constant 
of Universe expansion can be introduced:
\begin{equation}\label{relaxU}
 \tau_{\mathrm U}\simeq \sqrt{{{3c^2} \over { 8 \pi {G} \epsilon}}} =
32\,{\mu} \mbox{s}  \ \sqrt{{\epsilon_0}\over {\epsilon}},
\quad \epsilon_0  =1\,\frac{\mbox{GeV}}{\mbox{fm}^3}\,.
\end{equation}
The QCD energy scale $\epsilon_0=1$ GeV/fm$^3$
here used is not the total energy density in the Universe at the
time of hadronization. Aside of Quarks and gluons which comprise about 30 degrees of freedom 
of the visible matter energy density near to the hadronization condition there are 
furthermore 14.25   other  visible matter degrees of freedom (electrons, muons,
3 (left-handed) neutrinos and antineutrinos  and  photons). And there is dark matter.

\subsection{Dark matter at time of hadronization}
Potentially  relevant  is the influence of dark matter
on the dynamics of the QCD phase transition. 
Today, dark matter energy density (24\% of all) is
about 6 times as large as that of visible matter  (4.2\%);the balance ($72\pm5$\%) is 
the  so called dark `energy'. The  experimental evidence 
strongly favors dark energy in the form 
proposed by Einstein, i.e. a gravity repulsive $\Lambda$-term. If this is the case, the
presence of dark energy does not influence the quark Universe, since the energy 
density due to the Einstein $\Lambda$-term does not change with time, but has been always 
of the magnitude we see today. However, 
the visible energy scales  with $R^3$, or $R^4$, the faster scaling applies to 
radiation dominated era of the Universe. Since $R$ describs the growth of 
the Universe from hadronization era to present, without doubt the density of 
visible matter dominates the Einstein-like dark energy at hadronization
by an astronomical number of orders of magnitude.

The situation with dark matter is different. We do not know 
 how dark matter density extrapolates back in time to the QGP hadronization
era. We note that it is not possible 
that dark matter was a  negligible energy
component at time of QGP hadronization, since it is
 credited with the seeding of the (visible)
matter fluctuations, which ultimately were the cause 
of fast stellar and galactic structure development in the Universe. 
Thus one would be tempted to believe that dark matter was 
the dominant gravitational component in the early Universe,
also at the time of hadronization. The question is, if 
the time constants were accelerated by 1, 5, 10, or even 15 or more
 orders of magnitude, in which case the early 
Universe may have more resembled the fast exploding QGP
created in heavy ion collisions. 

We recall that the   visible matter  energy content  
has been converted into the background radiation and has been 
consumed  by the Universe expansion -- this is recognized by the large
entropy per baryon ratio. What we see today is a   tiny $10^{-9}$--$10^{-10}$ fraction 
of what was originally the visible energy content of the Universe. Thus 
both dark and visible energy content of the Universe may have changed
considerable since the quark-hadron epoch of the Universe.

We  close this discussion with a few  examples of 
 how  astro-particle models 
of dark matte,\cite{Bertone:2004pz}, and how these
may extrapolate back in time to hadronization epoch:
\begin{romanlist}[(ii)]
\item
So called `warm' dark matter candidate could be e.g.  a relatively light 
 sterile neutrino with mass of $m_{\nu s}<15$keV.\cite{Biermann:2006bu}  
Any such dark matter 
particle would have frozen out from the dense matter long before 
hadronization.   If we assume that their decay/annihilation is not
material   on time scale of 30$\mu$s,  we are 
considering an upper limit of what the influence of such dark matter
may be on the hadronizing Universe. 

We note that the Universe expansion  
`cools' the momentum $\vec p$ of dark matter $\vec p\to \vec p/R$,
while the energy density in the Universe is that of a radiation dominated Universe. 
Hence, as long as the ambient 
temperature $T\gg m_{\nu s} $  is higher than the warm dark matter particles, their  
 density  in the Universe is practically the same as that
of a thermally coupled particle. Therefore, 
at the time of QGP hadronization, such dark matter would be at most  
(i.e. if it does not decay or annihilate) another gravitating effectively massless particle,
contributing several (1,2, ?) degrees of freedom to 
the total count of about 45.  
Thus a sterile neutrino, or other warm dark matter would have marginal 
influence on dynamics of expanding quark-hadron Universe. 
\item
Dark matter particles in mass range of 
$m_d\simeq 1$--few MeV  could  
annihilate into $e^+e^-$  pairs.\cite{Boehm:2003bt}  
Like warm dark matter, at the time of QGP hadronization, this type of matter
would still be relativistic. The abundance can be expected to be 
in chemical equilibrium with its annihilation  products at the temperature 
well above the formation
threshold $e^+e^-\to 2m_d$.  We would have 
one (for the scalar model studied in depth,\cite{Boehm:2003bt}) 
or at most several, additional degrees of freedom, and again negligible 
impact on dynamics of the hadronizing Universe.

\item
Moving up in mass by a few orders in magnitude, we note that
dark matter could predominantly  consist of relatively very
heavy, as measured on current particle mass scale,
({\it e.g.\/}  super-symmetric) quasi-stable particles with $m_d\gg T_h$.
Such particles would make for cold dark matter at the time of QGP hadronization. 
Being cold at that time, as well as  now, and stable 
on scale of the Hubble-time, they  must have already frozen out, and   
could  not appreciably feed their energy into the expansion of the 
Universe in the  period following hadronization. The universe then 
and now would be filled with a dust of heavy invisible matter. This
type of dark matter is the most studied model, and decay rate limits
and annihilation rate limits are known for different types 
of particles.\cite{Bertone:2004pz} 

The 
worrisome thought is that  in principle, it is possible that  very massive 
dark matter decays or annihilates at a scale which would
deplete its number  by much more than a factor $10^{10}$ 
during the Universe expansion, and 
yet it would still be the dominant  matter form in the present day Universe. It suffices
to think of a family of dark matter particles with the lightest one being quasi-stable and 
contributing today to energy balance in the Universe. All we need to alter the picture of 
Universe expansion is that the 
 heavier particles are  stable on scale of 30$\mu$s, or even much shorter,
since the dark matter component shortens the natural time constant, 
see Eq.\,(\ref{relaxU}).

\end{romanlist}

 
\section*{Acknowledgements}
I would foremost like to thank  Walter Greiner for the great academic 
environment which I experienced  as a student 
in Frankfurt in the late 60's and 
early 70's.  Walter conveyed to his students his 
great love of physics, which accompanies us. 

I would further like to thank the FIAS (Frankfurt Institute for Advanced Studies) 
for very kind  hospitality which 
permitted participation at the ISHIP meeting in April 2006 
and this presentation.

 Research supported  
by a grant from: the U.S. Department of Energy  DE-FG02-04ER4131.
 

\end{document}